\documentclass[aps,pra,twocolumn,preprintnumbers,showpacs,superscriptaddress,amsmath,amssymb]{revtex4}
\usepackage{dcolumn}
\usepackage{bm}
\usepackage{amssymb}
\usepackage[german, english]{babel}
\usepackage{graphicx}
\usepackage{color}
\usepackage[letterpaper,total={7in,9.5in},top=0.75in,left=0.75in]{geometry}

\usepackage{array}

\begin{document}
\title{Radioluminescence and photoluminescence of Th:CaF$_2$ crystals}

\author{Simon Stellmer}
\affiliation{Institute for Atomic and Subatomic Physics, TU Wien, 1020 Vienna, Austria}
\affiliation{Vienna Center for Quantum Science and Technology, Vienna, Austria}
\author{Matthias Schreitl}
\affiliation{Institute for Atomic and Subatomic Physics, TU Wien, 1020 Vienna, Austria}
\author{Thorsten Schumm}
\affiliation{Institute for Atomic and Subatomic Physics, TU Wien, 1020 Vienna, Austria}

\date{\today}

% 06.30.Ft Time and frequency
% 78.20.-e Optical properties of bulk materials and thin films
% 78.47.jd Time resolved luminescence
% 78.60.Lc Optically stimulated luminescence

\begin{abstract}
We study thorium-doped CaF$_2$ crystals as a possible platform for optical spectroscopy of the $^{229}$Th nuclear isomer transition. We anticipate two major sources of background signal that might cover the nuclear spectroscopy signal: VUV-photoluminescence, caused by the probe light, and radioluminescence, caused by the radioactive decay of $^{229}$Th and its daughters. We find a rich photoluminescence spectrum at wavelengths above 260\,nm, and radioluminescence emission above 220\,nm. This is very promising, as fluorescence originating from the isomer transition, predicted at a wavelength shorter than 200\,nm, could be filtered spectrally from the crystal luminescence. Furthermore, we investigate the temperature-dependent decay time of the luminescence, as well as thermoluminescence properties. Our findings allow for an immediate optimization of spectroscopy protocols for both the initial search for the nuclear transition using synchrotron radiation, as well as future optical clock operation with narrow-linewidth lasers.
\end{abstract}

\maketitle

\section{Introduction}

Electronic transitions of valence electrons in atoms have typical energies of a few eV, whereas nuclear processes occur on the keV to MeV scale. This large gap in energy scales reflects in the fact that the realms of atomic and nuclear physics barely overlap, but there are a few exceptions.

One such exception is encountered in the nucleus of the isotope $^{229}$Th. This unique nucleus is believed to possess an extremely low-lying and long-lived excited state at an energy of a few eV, a property not found in any other known isotope \cite{Kroger1976fot,Reich1990eso,Helmer1994aes,Beck2007eso}. While direct evidence of the existence of this isomeric state is still pending \cite{Zhao2012oot,Jeet2015roa,Yamaguchi2015esf} and its energy has only been determined with large uncertainty, the fascinating possibility to manipulate nuclei by laser light has spurred a wealth of proposals for various applications.

The most prominent application of the $^{229}$Th isomer transition might be an optical clock based on this transition \cite{Peik2003nls,Campbell2012sin}. This clock could feature a quality factor of $Q=\nu/\Delta\nu \approx 10^{19}$, potentially outperforming today's best optical clocks \cite{Nicholson2015seo}. While such a clock might be highly immune to external perturbations, it would be very sensitive to variations of the fine-structure constant $\alpha$ and QCD parameters \cite{Flambaum2006eeo,Berengut2009pem,Rellergert2010cte}, constituting an exquisite probe of possible drifts in fundamental constants. In further quantum optics applications, the isomeric state has been proposed as a primer for the field of nuclear quantum optics \cite{Burvenich2006nqo,Liao2012ceo} and as a robust qubit for quantum information \cite{Das2013qie}. More generally, the unique case of $^{229}$Th might be the pioneer of gamma-ray lasers \cite{Tkalya2011pfa}. The prerequisite for all of these experiments is an unambiguous proof of the existence of the isomeric state, a measurement of its energy, and a demonstration of its optical addressability.

So far, most studies on $^{229}$Th employed high-resolution gamma spectroscopy \cite{Kroger1976fot,Reich1990eso,Helmer1994aes,Beck2007eso}. Differencing schemes were used to indirectly determine the energy of the isomeric state. The latest measurement places the excitation energy at 7.8(5)\,eV, corresponding to a wavelength of 159(10)\,nm in the vacuum-UV (VUV) range \cite{Beck2007eso,Beck2009ivf}. Systematic errors of this measurement might be underestimated \cite{Sakharov2010ote}. Additional evidence for the existence of an isomeric state has been obtained from collision experiments \cite{Burke1990aef}.

A number of experiments were performed to observe the VUV-photon emitted during the decay of the isomeric state. To date, all of these measurements generated false results \cite{Irwin1997ooe,Richardson1998upe} that were soon refuted \cite{Utter1999rot,Shaw1999sue}, or null \cite{Moore2004sfa,Inamura2009sfa,Jeet2015roa,Yamaguchi2015esf} results. The lifetime of the isomeric state is expected to be on the order of 1000\,s \cite{Helmer1994aes,Dykhne1998meo,Ruchowska2006nso}. Two experiments set out to measure the isomer lifetime via alpha spectroscopy \cite{Kikunaga2009hle} and through gamma decay \cite{Browne2001sfd}, but found no signal. A recent experiment claims observation of the VUV photon with an isomer lifetime of 6(1)\,hours \cite{Zhao2012oot}, but is highly disputed \cite{Peik2013coo}.

There exist a large number of strategies to populate the isomeric state. So-called ``indirect'' pathways include the alpha decay $^{233}\textrm{U} \rightarrow {}^{229m}\textrm{Th}$ \cite{Zhao2012oot,Wense2013tad}, excitation via higher-lying nuclear states (e.g.~at 29.19 keV) \cite{Tkalya2000dot}, electron bridge processes \cite{Porsev2010eot}, and light fusion reactions \cite{Sonnenschein2012tsf}. Direct optical excitation of the isomeric state via a photon with the correct wavelength has not yet been successful, nor has the wavelength of the transition been determined with sufficient precision to commence spectroscopy with narrow-line lasers. Spectroscopy with synchrotron radiation may fill this apparent gap \cite{Jeet2015roa}.

Optical spectroscopy of the $^{229}$Th isomeric transition requires a platform to securely hold the nucleus for a sufficiently long time in a recoil-free environment. Radiation with a wavelength shorter than 196\,nm ionizes the neutral Th atom (first ionization energy 6.32\,eV), experiments should therefore employ positively charged Th ions for spectroscopy in the VUV. The extreme scarcity of the isotope $^{229}$Th, related to its half-life of $\tau = 7932\,a$ \cite{Kikunaga2011dot}, and the minuscule cross section for optical excitation \cite{Kazakov2012poa}, place further demands on the spectroscopy scheme. Two platforms have been proposed as hosts of $^{229}$Th ions: ion traps \cite{Peik2003nls,Campbell2011wco,Herrera2014elo} and VUV-transparent crystals \cite{Peik2003nls,Hehlen2013oso}.

\begin{figure}
\includegraphics[width=\columnwidth]{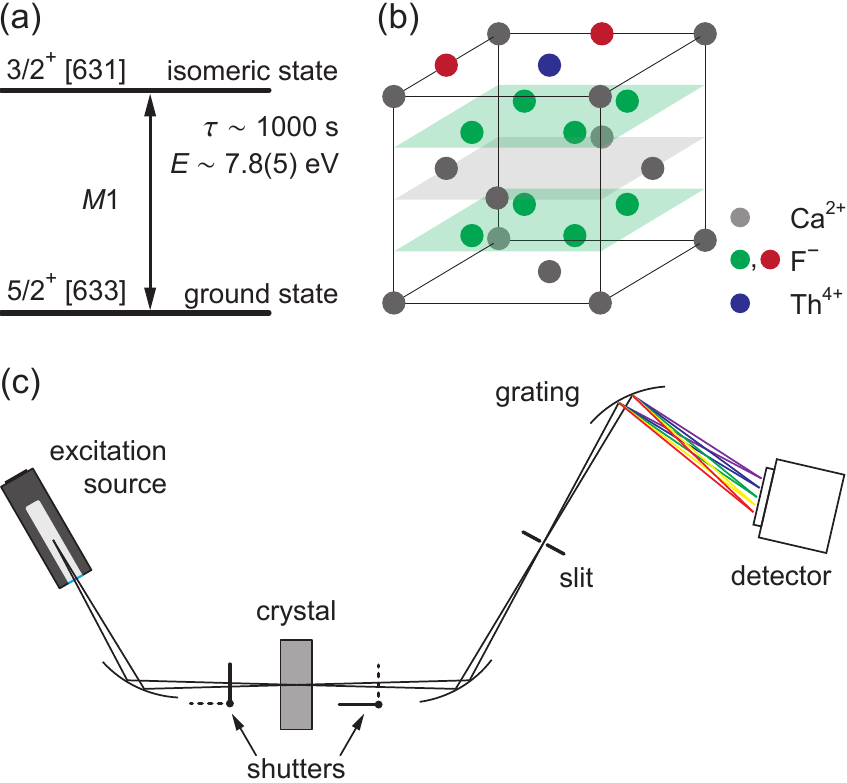}
\caption{(a) The nuclear two-level system in $^{229}$Th. (b) Lattice structure of Th:CaF$_2$, where a Th$^{4+}$ ion replaces a Ca$^{2+}$ ion and the additional charge is compensated by two F$^-$ interstitials at 90$^{\circ}$ angle. (c) Experimental set-up:~a spectrometer is used to measure the crystal's luminescence spectrum. A deuterium lamp serves for both the excitation of the crystal and the calibration of the spectrometer. A manual actuator allows to place different probes into the central region. The instrumental resolution is about 1 nm.}
\label{fig:fig1}
\end{figure}

CaF$_2$ is a promising candidate for such a host crystal \cite{Karpeshin2007iot,Kazakov2012poa}. Its fairly simple lattice has a band-gap of about 12\,eV, which ensures optical transparency down to 120 nm. CaF$_2$ readily accepts thorium as a dopant \cite{Schreitl2015PhD}. Lattice structure calculations show that Th$^{4+}$ ions replace Ca$^{2+}$ ions, where the additional charges are compensated for by flourine interstitials \cite{Dessovic2014tdc}; see Fig.~\ref{fig:fig1}(b). Doping with Th may reduce the band-gap of CaF$_2$ by a few percent \cite{Dessovic2014tdc}, but Th:CaF$_2$ crystals with doping concentration below $10^{-4}$ remain transparent down to 125\,nm \cite{Schreitl2015PhD}. Broadening mechanisms caused by the lattice environment \cite{Kazakov2012poa} might limit the ultimate performance of clock operation with this platform, but the ability to place more than $10^{15}$ nuclei into a volume of 1\,cm$^{3}$ is a tremendous advantage for the initial optical search for the transition. This \emph{solid state} approach thus enables optical M\"ossbauer spectroscopy and provides a very simple, robust, and secure way to store the $^{229}$Th nuclei. Figure~\ref{fig:fig1} (a) summarizes the relevant properties of the isomer transition.

The long lifetime of the isomeric state can only be exploited if radiative $M1$ decay into the nuclear ground state is the dominant pathway of decay. Thus, competing pathways such as fast non-radiative relaxation and internal conversion need to be suppressed \cite{Tkalya2000dot,Karpeshin2007iot}. This requires the $^{229}$Th nucleus to be located at a well-defined lattice site, avoiding any coupling to the crystal's electronic states. The actual microscopic environment of Th ions in the crystal is still to be explored.

Pure CaF$_2$ is widely used for UV optics, and numerous studies have investigated scintillation properties in response to irradiation with hard X-rays, as well as resilience against intense pulsed VUV light (e.g.~in Refs.~\cite{Williams1976trs,Mikhailik2006spo,Rodnyi1997ppi}). These studies only covered parameter regimes far away from the one relevant for the work presented here.

In this paper, we study the suitability of $^{229}$Th:CaF$_2$ crystals for optical spectroscopy of the $^{229}$Th nuclear transition. We suspect that the nuclear signal might be masked by two types of background:~VUV-photoluminescence of the crystal, caused by the interrogation light (Sec.~\ref{sec:photoluminescence}), and radioluminescence, caused by radioactive decay of $^{229}$Th (Sec.~\ref{sec:radioluminescence}). We employ VUV light to induce photoluminescence, and take advantage of the radioactivity of thorium nuclei doped into the crystal as an intrinsic source of radioluminescence. A spectrometer with nm-resolution is used to measure the emission spectrum of both kinds of luminescence. In time-resolved studies, we use photo-multiplier tubes (PMTs) to measure the duration of crystal scintillation.

\section{Experimental set-up}

\subsection{CaF$_2$ crystals}

We operate an in-house furnace to grow CaF$_2$ single crystals by means of the vertical gradient freeze technique \cite{Schreitl2015PhD}. The crystals have a volume of 7\,cm$^3$ and are cut and polished to disks of typically 5\,mm thickness. For the studies presented here, we employ pure CaF$_2$ crystals, $^{232}$Th:CaF$_2$ crystals, and  $^{229}$Th:CaF$_2$ crystals. The isotope $^{232}$Th (half-life $\tau=1.4 \times 10^{10}\,$a) is readily available and chemically identical to $^{229}$Th, and can therefore be used as a substitute. The doping concentration (fraction of Ca$^{2+}$ ions replaced) can be varied from $10^{-8}$ to values as high as 1\%. The limited availability of the isotope $^{229}$Th restricts the concentration in our crystals to $1.6 \times 10^{-8}$, corresponding to a density of $4 \times 10^{14}\,{\rm cm}^{-3}$. As a result of the chemical processing of $^{229}$Th, the $^{229}$Th:CaF$_2$ crystals also contain $^{232}$Th at a concentration of $10^{-4}$.

Pure CaF$_2$ is commonly used for UV optics, therefore some of its scintillation properties have already been studied. Such studies found that self-trapped excitons (STEs) are the dominant type of excitation created by various kinds of radiation \cite{Williams1976trs,Mikhailik2006spo,Rodnyi1997ppi}. Any form of ionizing radiation absorbed by the crystal will create highly energetic electron-hole pairs, which quickly thermalize to form a myriad of various excitations with energies below the band gap. These may transfer their energy to various kinds of luminescence centers, which eventually decay under emission of photons. The temperature-dependent relaxation timescales range from ns to ms. A particular type of defect are STEs, which can heal by emission of UV photons at a wavelength of around 300\,nm. The ratio of non-radiative (via lattice phonons) to radiative relaxation is temperature-dependent \cite{Rodnyi1997ppi}.

\subsection{Excitation}

Various sources of optical excitation are available, such as deuterium and noble gas lamps, excimer lasers, synchrotron radiation, and solid state lasers. These light sources greatly differ in spectral photon flux, spectral width, and tunability. In this work, we will focus on deuterium lamps and excimer lasers.

\subsection{Detection}

Commonly used strategies for the detection of UV photons differ in their spectral and temporal resolution.

Photo-multiplier tubes (PMTs) offer single-photon detection and have excellent timing resolution down to well below 10\,ns. To optimize detection efficiency at the desired wavelength, various materials can be chosen for the photocathode. Cs-I cathodes are sensitive between 115\,nm (the transmission edge of the MgF$_2$ window) and 190\,nm, diamond cathodes operate up to 220\,nm, and Cs-Te cathodes up to 320\,nm. The peak quantum efficiency of VUV PMTs is of order 10\%. The sensitivity at higher wavelengths, however, is not zero, but about 1/1000 of the peak sensitivity. PMTs thus offer a certain degree of spectral filtering with a bandwidth of order 100\,nm, which might be improved by UV filters. Adding a spectrometer to the detection system offers spectral resolution down to below 0.1\,nm, however the light throughput is fairly small, reducing the signal by typically five orders of magnitude. The reduced S/N often leads to long integration times and poor temporal resolution.

Studies in the VUV range are hampered by the fact that molecular oxygen absorbs light at wavelengths below 180\,nm, therefore all experiments take place in an oxygen-depleted environment. We operate two vacuum systems:\,one is equipped with a VUV spectrometer and a deuterium lamp, the other one holds a set of different PMTs and an excimer laser as light source.

\subsubsection{Spectrometer set-up}
Studies that require spectral resolution are performed in a UV spectrometer (McPherson model 234/302); see Fig.~\ref{fig:fig1}(c). The crystal is imaged onto a slit with a typical width of $600\,\mu$m, and the slit in turn is imaged onto the sensor of a CCD camera by a concave holographic grating. We used various gratings with groove densities between 300 and 2400\,g/mm, but most measurements of this work were performed with a 600\,g/mm grating (blaze angle optimized for 150\,nm). The total efficiency of the spectrometer is $2 \times 10^{-7}$, largely limited by the small solid angle and a grating efficiency of a few percent. We determine the instrumental resolution by setting the grating to the blaze angle, thereby imaging the spectrometer's slit directly onto the camera. The best resolution obtained for minimum slit widths is 0.5\,nm.

The camera is an ANDOR Newton 940 model with a pixel size of 13.5\,$\mu$m, the chip is cooled to $-95\,^{\circ}$C. The chamber is pumped down to a pressure of $2 \times 10^{-6}$\,mbar to avoid spurious signals from atoms and molecules in the residual gas \cite{Utter1999rot,Shaw1999sue}. The entire chamber is housed in a lead shield of 20\,mm thickness to protect the CCD camera from ambient gamma radiation of up to 1\,MeV energy. This measure reduces the rate of ``cosmic'' background events by a factor of at least 4, consistent with literature values. Given the long exposure times and large binning areas used in most measurements, however, the signal background is still dominated by ``cosmic'' events, both massive high-energy particles and gammas. We develop a reliable protocol to detect and remove such events in our data sets.

The spectrometer is calibrated by a Hamamatsu L1835 deuterium lamp, which is also used for illumination of the crystal. The lamp has a strong characteristic emission spectrum between 120 and 170\,nm, the light intensity on the crystal is a few W/cm$^2$ with a spot size of 3\,mm. The absolute calibration and its reproducibility are better than 1\,nm.

\subsubsection{PMT set-up}

The time evolution of luminescence signals is investigated with PMTs. Measurements of the radioluminescence are performed in a dedicated vacuum chamber evacuated to $10^{-5}$\,mbar. We employ PMTs with Cs-I, diamond, and Cs-Te photocathodes (Hamamatsu models R6835, R7639, and R6836) in head-on and side-on configurations. The PMTs are cooled to $0\,^{\circ}$C. Typical settings are a high voltage of $-1500$\,V and a signal threshold of $-7.8\,$mV, resulting in a dark count rate of 5\,Hz. The voltage divider circuitry is placed inside the vacuum, and the signals are counted by a Becker + Hickl card without prior amplification. The FWHM of a signal is below 2\,ns, and we allow for a hold-off time of 5\,ns.

The crystal temperature can be controlled between 0 and $160\,^{\circ}$C using peltier elements and heating wires, and the distance between crystal and PMT can be varied to adjust the absolute count rate.

A 157-nm F$_2$ excimer laser is attached to this chamber to induce photoluminescence in the crystal. The pulse energy and duration are 1\,mJ and 7\,ns, respectively, the repetition rate is set to 100\,Hz, and the size of the beam is $8\times3$\,mm$^2$. Mechanical shutters are used to protect the PMT from direct exposure to the excitation light.

\section{VUV-photoluminescence}
\label{sec:photoluminescence}

\subsection{Optical spectrum}

\begin{figure}
\includegraphics[width=\columnwidth]{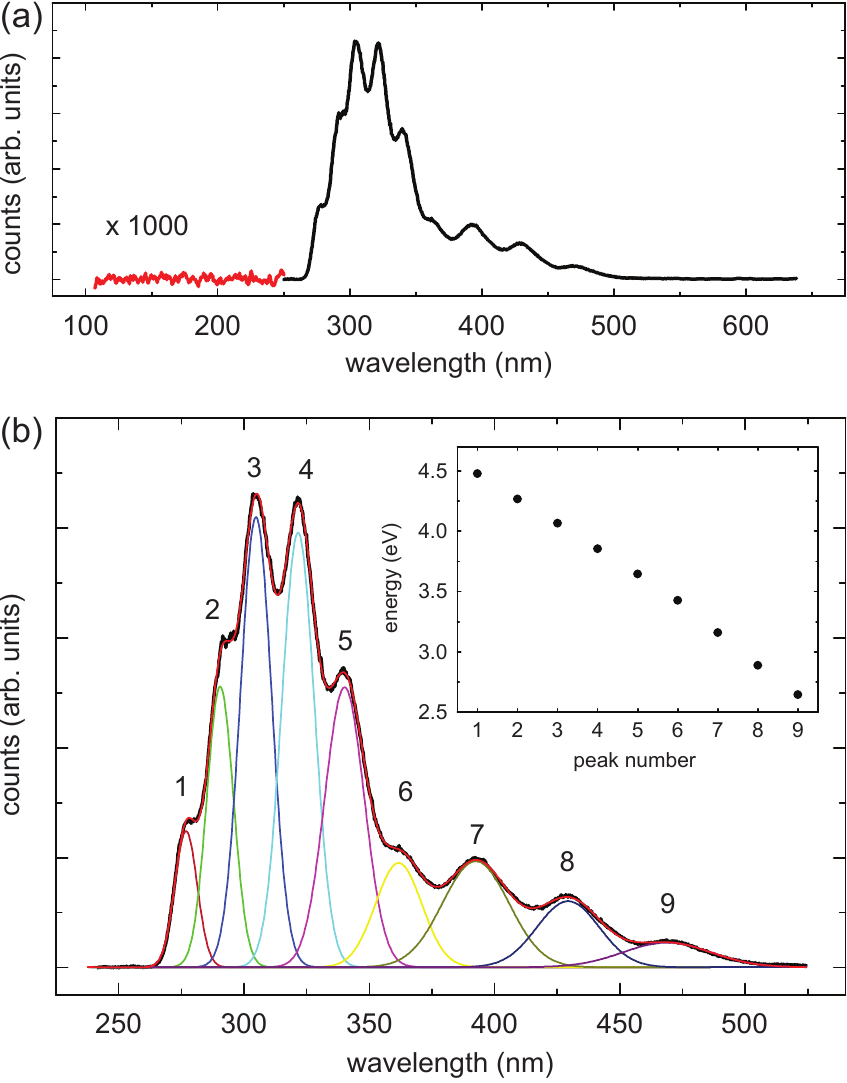}
\caption{VUV photoluminescence spectrum of CaF$_2$. (a) The spectrum shows characteristic lines between 260 and 500\,nm. The data below 250\,nm is magnified by a factor of 1000 to demonstrate the absence of crystal emission in this region. (b) Zoom into the self-trapped exciton (STE) feature, where the data (black dots) is fitted by nine Gaussian peaks of variable position, amplitude, and width (colored lines). The positions of the Gaussian peaks, expressed in units of eV, are plotted in the inset. The instrumental resolution is 0.5\,nm.}
\label{fig:fig2}
\end{figure}

We begin our studies by an investigation of the VUV-light induced luminescence. A deuterium lamp is used to simulate a synchrotron or a future tunable and narrow-band spectroscopy light source. A home-grown $^{232}$Th:CaF$_2$ crystal is illuminated for 1\,s and subsequently exposed to a CCD camera via a spectrometer for one second. This alternating cycle of crystal illumination and detection is repeated for an hour to improve the signal-to-noise ratio. The measured spectrum shows distinct peaks between 260 and 500\,nm, interpreted as STE emission, but no discernable features below 260\,nm; see Fig.~\ref{fig:fig2}(a).

The STE spectrum consists of nine overlapping lines, which can be approximated by Gaussian functions. In the inset of Fig.~\ref{fig:fig2}(b), we plot the positions of the lines, where the wavelength $\lambda$ has been converted into photon energy via $E = h c/\lambda$. The dependence of photon energy on peak number is almost linear, suggesting that each line represents an oscillator state in the harmonic STE potential. The width of the lines is between 5 and 10\,nm (substantially larger than the instrumental resolution) and decreases linearly with increasing energy. %Maybe a calibration of the spectrometer needs a quadratic term, or it's just not a harmonic potential.

The position of the lines is independent of crystal temperature, wait time after illumination, light intensity, and emission spectrum of the deuterium lamp. Their relative amplitudes, however, do change with wait time after illumination, and spectrum of the excitation source. Notably, we varied the illumination/exposure time between 250\,ms and 10 hours and found only little change to the overall shape of the spectrum.

We do observe a weak temperature dependence of the emission strength. Between 20 and $110\,^{\circ}$C, the emission increases by about 60\% in a near-linear fashion, but the relative amplitudes of the individual lines are unchanged.

It should be noted that STEs are a well-known type of excitation in rare-earth halides \cite{Rodnyi1997ppi}. The VUV light might not excite the STEs directly, but may induce various kinds of other crystal excitations, which quickly relax into STEs. The magnitude of the excitonic luminescence appears to depend on the purity of the crystal, and therefore on the details of the crystal growing process. For our home-grown crystals, the amplitude of luminescence emission varies by about a factor of five within a large set of home-grown specimens, irrespective of the doping concentration. We probe various commercial CaF$_2$ crystals (Hellma/Schott Lithotec and Korth) and find their VUV-photoluminescence to be a factor of at least $10^4$ smaller than for our home-grown crystals. A $^{232}$Th:CaF$_2$ crystal produced by the Institut f\"ur Kristallzucht (IKZ, Berlin, Germany) using the Czochralski method shows negligible luminescence. We thus conclude that the emission amplitude depends only on the quality of the CaF$_2$ production process, with no discernable influence of the thorium doping concentration.

An absolute quantification of the crystal's photoluminescence in response to prior illumination with a well-defined light source is beyond the scope of this paper. We estimate that our home-grown 10\,mm thick CaF$_2$ crystals, illuminated by a deuterium lamp with 200\,mW optical power for 30 minutes, radiate $10^8$ photons/s one minute after the end of illumination.

\subsection{Timescales}

\begin{figure}
\includegraphics[width=\columnwidth]{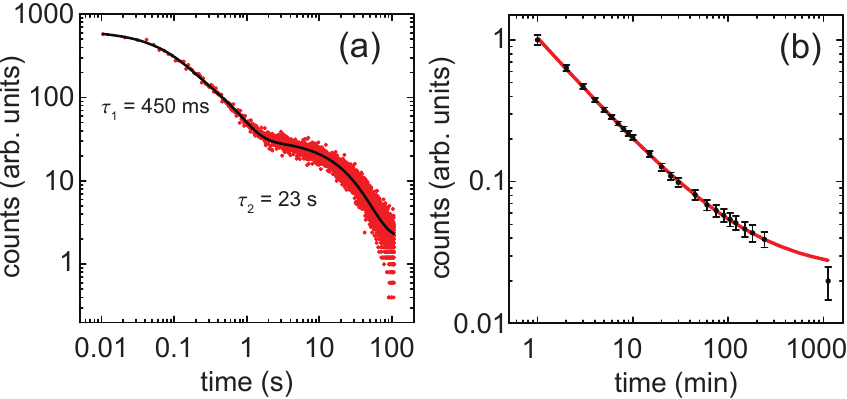}
\caption{Time-dependent decay of the VUV-light induced luminescence, measured separately for timescales up to 100\,s (a), and up to 20 hours (b). The data is fit by a double-exponential decay for short wait times, and by a power law for longer times.}
\label{fig:fig3}
\end{figure}

The decay of luminescence in CaF$_2$ on ns- and $\mu$s-timescales has already been studied extensively \cite{Williams1976trs}. In future spectroscopy and optical clock experiments, however, mechanical shutters have to be opened and closed in between crystal illumination and fluorescence measurement, introducing a delay of about 10\,ms. Consequently, we limit our studies to timescales longer than 10\,ms.

An excimer laser at 157\,nm is used to illuminate the crystal for one minute, and the subsequent luminescence detection is performed with a Cs-Te PMT. For detection times below one minute, we find a double-exponential decay with time constants $\tau_1 = 450\,$ms and $\tau_2 = 23\,$s; see Fig.~\ref{fig:fig3} (a). These values change slightly with varying experimental parameters, but the significant difference in timescales prevails. The amplitude of the faster decay is at least 10-times larger compared to the slower branch. For times in excess of one minute, we observe the familiar power-law decay of luminescence \cite{Jonscher1984ttd,Huntley2006aeo}. This decay that can be traced for times as long as a full day; see Fig.~\ref{fig:fig3} (b). We fit the data by $I(t)/I_0 = c + t^{-k}$, where $c$ is an offset related to radioluminescence, and obtain an exponent $k=0.75(2)$ at room temperature.

To conclude, we find that the STE spectrum in CaF$_2$ has an abrupt lower edge at 260\,nm, and we observe no emission in the relevant spectral region around 160\,nm. Choosing PMTs with Cs-I or diamond photocathodes ensures maximum detection efficiency in the desired wavelength range and suppresses the luminescence background by at least three orders of magnitude. In addition, a vast fraction of the luminescence decays on timescales much shorter than the expected isomer lifetime. These findings suggest that both temporal and spectral filtering can be used to discriminate the nuclear spectroscopy signal from crystal luminescence.

\section{Radioluminescence}
\label{sec:radioluminescence}

The alpha decay of $^{229}$Th in the crystal is a violent process:~the alpha particle and the remnant $^{225}$Ra nucleus obtain kinetic energies of 5.1\,MeV and 90\,keV, respectively. These fragments travel through the crystal lattice structure with ranges of about 30\,000 and 30 lattice constants, respectively, leaving behind a track of defects. From an energy point of view, each alpha decay releases enough energy to create $10^6$ photons. While most of the released kinetic energy $Q_{\alpha}$ is eventually converted into phonons, some of these defects relax via emission of photons, known as scintillation. As we will show later, each alpha decay in CaF$_2$ generates $10(2)\times 10^3$ photons in the UV range. In addition, all the daughter products of $^{229}$Th down to $^{209}$Bi are short-lived (half-lives between $3.7\,\mu$s and 15 days), such that each $^{229}$Th decay is followed by a chain of four alpha and three beta decays.

Radioluminescence thus poses a considerable background for both the initial search for the isomeric transition, as well as future operation of an optical clock, where the background inevitably scales linearly with the nuclear isomer signal. In the following, we will characterize the radioluminescence with respect to its spectrum and its dependence on time and temperature.

\subsection{Optical spectrum}

\begin{figure}
\includegraphics[width=\columnwidth]{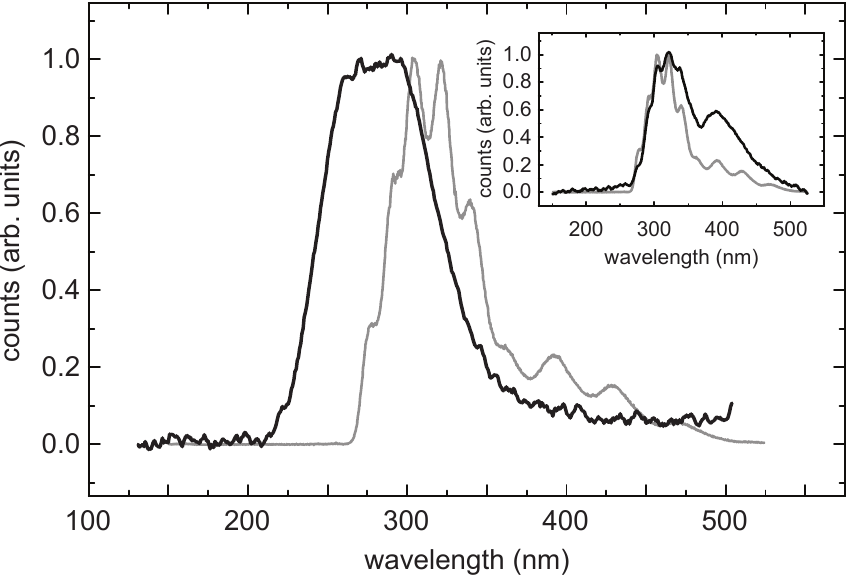}
\caption{Radioluminescence spectrum of a $^{229}$Th:CaF$_2$ crystal (black line), compared to a scaled photoluminescence spectrum (grey line). The instrumental resolution is 15\,nm. Inset: Heating the crystal to $150\,^{\circ}$C anneals permanent defects under an emission of a spectrum very similar to photoluminescence.}
\label{fig:fig4}
\end{figure}

We place a $^{229}$Th:CaF$_2$ crystal (thickness 20\,mm, doping concentration $1.6 \times 10^{-8}$, activity 5\,kBq) into the spectrometer and integrate the emission spectrum for 100 hours. The radioluminescence spectrum is shown in Fig.~\ref{fig:fig4}; spectra obtained with highly-doped $^{232}$Th:CaF$_2$ crystals are identical. Importantly, no emission is observed for wavelengths below 220\,nm. These spectra show pronounced emission between 230 and 400\,nm and differ significantly from the ones induced by VUV light. The crystal was heated to $300\,^{\circ}$C prior to the measurement to remove emission related to thermoluminescence (Sec.~\ref{sec:thermoluminescence}).

\subsection{Timescales}

\begin{figure}
\includegraphics[width=\columnwidth]{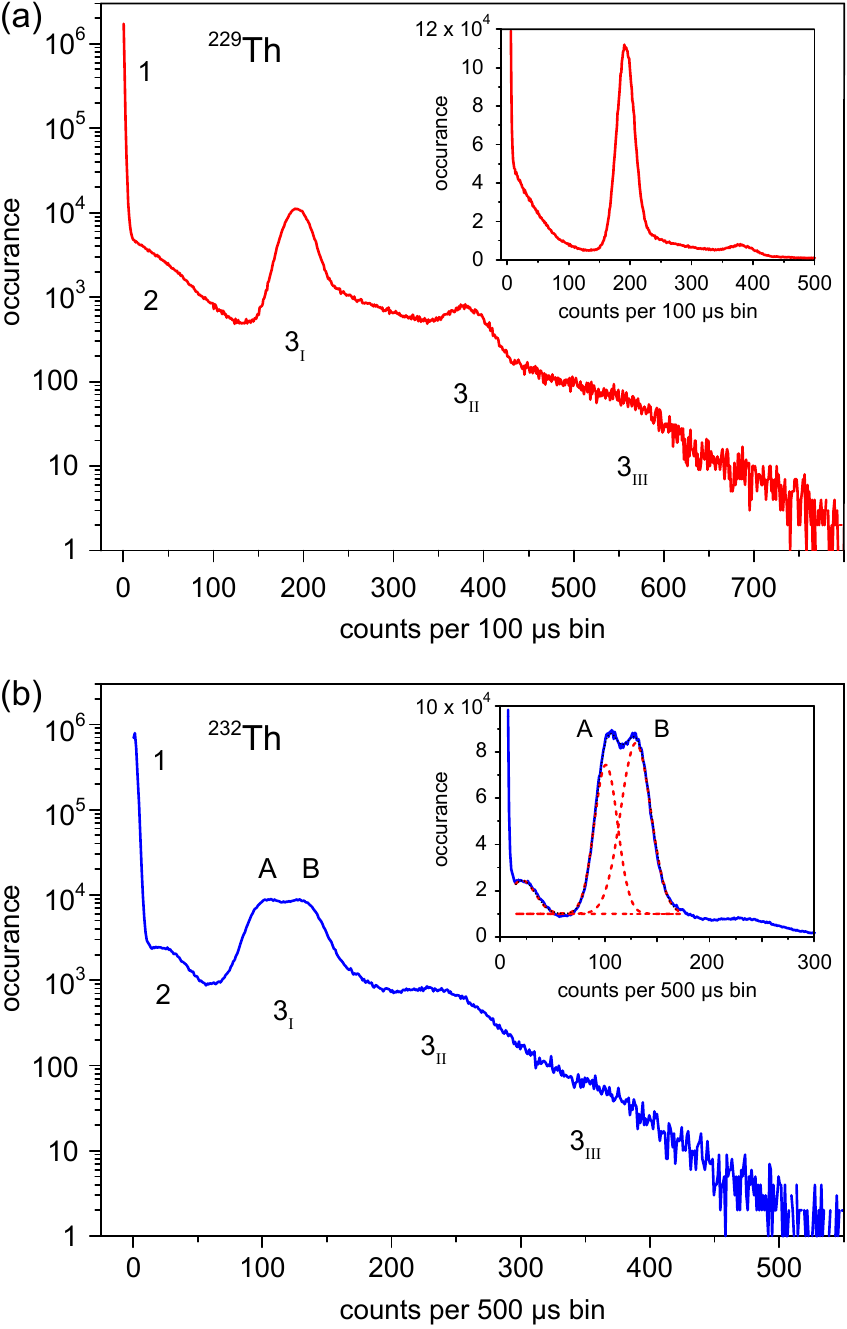}
\caption{Bursts of radioluminescence induced by alpha decay. (a) Single photons emitted by a $^{232}$Th:CaF$_2$ crystal are detected by a PMT and counted in bins of $100\,\mu$s width. The histogram shows a summation of $3 \times 10^6$ bins, the numbers label relevant features discussed in the text. A section of the data is also plotted on a linear scale (inset). (b) Similar measurement for a $^{232}$Th-doped crystal. The feature labelled ``B'' points to a $^{229}$Th contamination of the crystal; see the text for details. The dashed red lines in the inset are Gaussian fits to the three center peaks.}
\label{fig:fig5}
\end{figure}

We use a Cs-Te PMT (sensitivity range 115 to 320\,nm) to record the emission of a $^{229}$Th:CaF$_2$ crystal (activity 1\,kBq) at the single-photon level. The bin width is set to $100\,\mu$s, a factor of 10 shorter than the average time difference between two successive alpha events. The total detection efficiency is about 2.5\%. A typical histogram of counts per bin, using 5 minutes of integration, is shown in Fig.~\ref{fig:fig5} (a).

We observe five clearly distinguished features. The vast majority of bins contain zero or very few counts, forming the Poissonian distribution labelled ``1'' in Fig.~\ref{fig:fig5} (a). Strikingly, about 13\% of all bins make up feature $3_{\mathrm{I}}$, each containing about 200 counts. We interpret this feature as a flash of photons, released in succession of an alpha decay inside the crystal, with a duration much shorter than $100\,\mu$s. The rate of such events matches the $^{229}$Th activity inferred by neutron activation analysis and gamma spectroscopy. Consequently, bins containing two and three of such events constitute features $3_{\mathrm{II}}$ and $3_{\mathrm{III}}$ in the figure. Feature 2 is a signature of the beta decay of $^{229}$Th daughters and possibly contaminations in the crystal.

Next, we sandwich the crystal in between two identical PMTs. We observe the appearance of ``bursts'' always in coincidence on both detectors, again supporting the hypothesis that such flashes are related to alpha decay in the crystal. We calculate that each decay gives birth to $10(2)\times 10^3$ photons, thus about 1\% of $Q_{\alpha}$ is radiated via photons.

We then exchange the $^{229}$Th:CaF$_2$ crystal for a $^{232}$Th:CaF$_2$ specimen. The doping concentration of 0.8\% is nearly 6 orders of magnitude higher to account for the long half-life of $^{232}$Th ($\tau = 1.4 \times 10^{10}$\,a). We use a longer binning time of $500\,\mu$s; the obtained histogram is shown in Fig.~\ref{fig:fig5} (b). The fact that the main peak has shifted to a lower value of counts is explained by the reduced optical transmission properties of this specific crystal. The upshift of feature 2 can be explained as a signature of the beta decay of $^{228}$Ac in the $^{232}$Th decay chain:~a comparably large amount of kinetic energy is released in this decay ($Q_{\beta^-} = 2.124$\,MeV). Interestingly, two separate peaks A and B are clearly visible, and we speculate that they correspond to two types of alpha decay with different values of $Q_{\alpha}$. Fitting simple Gaussian distributions to the dominant peaks (dashed red lines in the inset of Fig.~\ref{fig:fig5} (b)), we obtain a difference in released photon number of 27.8(6)\% between peaks A and B. Following the Geiger-Nuttall law, we assign feature A (lower $Q_{\alpha}$) to the very long-lived $^{232}$Th isotope. The $^{229}$Th value of $Q_{\alpha}$ is 26.6\% larger compared to $^{232}$Th \cite{nudat}, very close to the difference in observed photon numbers. We thus identify feature B with a crystal contamination of $^{229}$Th at a level of 0.8(1)\,ppm relative to the $^{232}$Th content.

We repeat the same experiment with a $^{232}$Th:CaF$_2$ specimen (doping concentration $2 \times 10^{-4}$) produced at IKZ. This crystal shows the very same radioluminescence signature as our home-grown $^{232}$Th:CaF$_2$, albeit different contaminations. Note that the IKZ crystal showed negligible signs of photoluminescence. This lets us to believe that the observed radioluminescence is a generic scintillation feature of CaF$_2$ and does not depend on the details of the crystal growing process.

\begin{figure}
\includegraphics[width=\columnwidth]{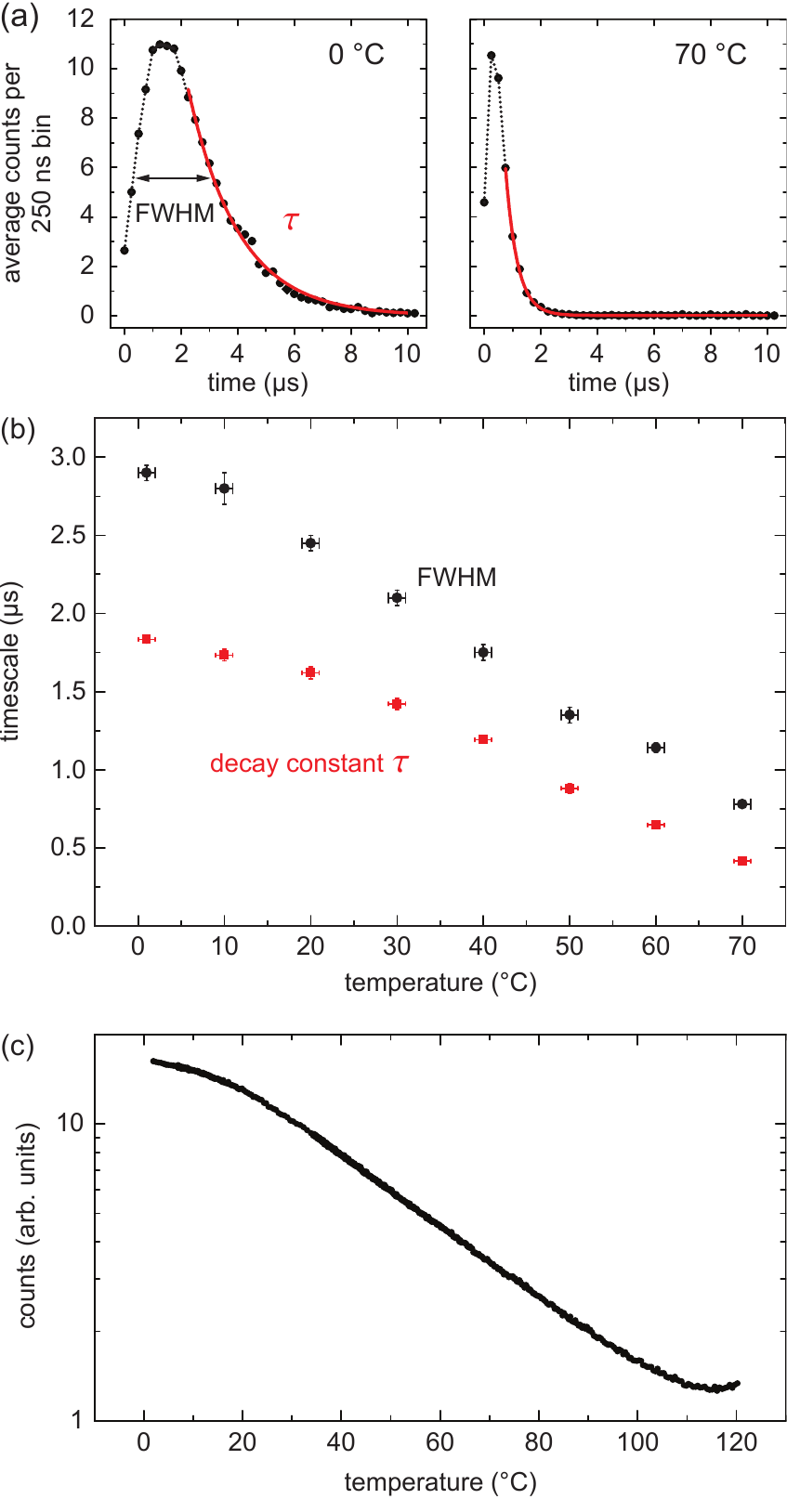}
\caption{Temperature dependence of radioluminescence in $^{229}$Th:CaF$_2$. (a) Time evolution of a burst of photons initiated by an alpha decay, shown for two different temperatures. An exponential decay (red line) is fit to the tail of the distribution; the dotted line connects data points (black dots) to guide to the eye. (b) The duration of a burst depends on temperature, quantified here by the full width at half maximum (FWHM) and the decay time $\tau$ of the burst. (c) The integrated photon emission changes by a factor of 10 between 0 and $100\,^{\circ}$C.}
\label{fig:fig6}
\end{figure}

We found that the characteristic ``flashes'' are substantially shorter than $100\,\mu$s. To measure their duration more precisely, we reduce the detector bin width to 250\,ns and record the emission of a $^{229}$Th-doped crystal. We find that each flash lasts on the order of $1\,\mu$s, the average time interval between flashes being about 1\,ms. From the PMT's time trace, we extract sequences that contain about 200 counts within a few $\mu$s, thus fall into the prominent peak 3 in Fig.~\ref{fig:fig5} (a). To reduce the shot noise, we average over 100 of such flashes, where the first non-zero bin is taken as the onset of the flash. A typical time evolution curve is shown in Fig.~\ref{fig:fig6} (a).

In view of the coarse timing resolution, as well as the lack of a profound model to explain the time evolution, we will restrict ourselves to a semi-quantitative analysis. A simple exponential decay fits the tail of the curve surprisingly well, yielding a value of 1.6(1)\,$\mu$s for a crystal at room temperature. The FWHM of the distribution gives a similar value, 2.3(1)\,$\mu$s. Substituting $^{229}$Th by $^{232}$Th in the crystal yields the same result. Changing key parameters of the PMT set-up, such as the hold-off time and absolute count rate, we find only mild changes. Note that even for the highest count rates of about 50\,MHz during a flash, the average interval between counts is still larger than the PMT's signal width (2\,ns) and the typical holdoff-time (5\,ns).

The near-unit efficiency in detecting alpha decays in the crystal could be turned into a powerful tool to reduce the radioluminescence background of $^{229}$Th nuclear spectroscopy measurements. A primary PMT, most sensitive around 160\,nm, would be used to detect the sought-after emission of the isomeric state. This PMT would have a non-negligible relative sensitivity of $10^{-3}$ for radioluminescence photons. An ancilla PMT, sensitive around 300\,nm, would be used to detect alpha decays with high efficiency, thus recording the time stamp of every event. When post-processing of the data, any signal registered by the primary PMT during a $\mu$s time window around the alpha event would be excluded. The same could be performed already on-line with suitable gating scheme.

\subsection{Temperature-dependence}

As a next step, we measure the duration of the flashes at different crystal temperatures; see Fig.~\ref{fig:fig6} (b). Between 0 and $70\,^{\circ}$C, the two extracted measures of the flash duration decrease by a factor of about 4 in a near-linear fashion. This speed-up is identical for $^{229}$Th- and $^{232}$Th-doped crystals.

Not only the duration of each flash, but also the number of contained photons is strongly temperature-dependent: the ratio of radiative to non-radiative relaxation of STE defects changes drastically around room temperature \cite{Rodnyi1997ppi}. Recording histograms as in Fig.~\ref{fig:fig5}, we find that the prominent peak shifts to the left for increased temperature:~for every alpha decay, fewer photons are emitted. More quantitatively, we measure the total emission of a crystal by integrating over time intervals of 10\,s, thereby detecting not only prompt emission following alpha decays, but also relaxation of long-lived defects and beta decays. Such a curve, taken with a Cs-Te PMT right after annealing a $^{229}$Th:CaF$_2$ crystal at $300\,^{\circ}$C, is shown in Fig.~\ref{fig:fig6} (c). Between 30 and $100\,^{\circ}$C, the photon emission decreases exponentially with temperature, dividing in half every 24\,K. The apparent increase above $110\,^{\circ}$C is caused by black-body radiation emitted by the crystal heater.

We perform the same measurement using a PMT with a diamond photocathode, for which the spectral sensitivity has a plateau up to 170\,nm and gradually decreases towards higher wavelengths. The signal amplitude is drastically reduced, yet we find the same temperature dependence. This indicates that Cherenkov radiation between 115 and 220\,nm, which we do not assume to be temperature-dependent, comprises less than 1\% of the photons emitted as a result of radioactivity. This agrees with the spectrum shown in Fig.~\ref{fig:fig4}, where negligible Cherenkov radiation was observed.

This finding is again important for future $^{229}$Th spectroscopy measurements: Mild heating of the crystal to around $100\,^{\circ}$C reduces its radioluminescence by a factor of 10 from its room temperature value. In addition, the duration of characteristic flashes decreases dramatically, which in turn reduces the veto time imposed onto the primary detector to reject unwanted counts during radioluminescence flashes.

\subsection{Thermoluminescence}
\label{sec:thermoluminescence}

Ionizing radiation can create semi-permanent defects in doped CaF$_2$, exhibiting lifetimes of several weeks or months. Such excitations, often related to contaminations or crystal defects, can be healed by heating of the crystal, allowing excited electrons to leave their traps and relax into the ground state by photon emission. This behavior is widely used in thermoluminescent dosimeters (TLDs) based on CaF$_2$.

We allow a $^{229}$Th:CaF$_2$ crystal to accumulate long-lived defects for many weeks. Heating the crystal to $150\,^{\circ}$C frees an enormous amount of photons. The spectrum, shown in the inset of Fig.~\ref{fig:fig4}, is markedly different from the one obtained with ``prompt'' photons. We observe a series of overlapping lines, where the position of the individual lines (but not their relative amplitudes) coincides perfectly with the photoluminescence spectrum.

Next, we use the single photon counting capability of PMTs to access the absolute amount of photons radiated upon heating. Measuring an ordinary glow curve, we estimate that for each alpha decay, a few $10^4$ semi-permanent defects are created in the crystal. These defects have a lifetime of months, and slowly relax under emission of single uncorrelated photons. These photons (and not the PMT dark counts) are the origin for the feature labelled ``1'' in Fig.~\ref{fig:fig5} (a). Thus, for every alpha event, there are more semi-permanent defects created than prompt photons released.

These ``delayed'' photons pose an unpleasant background to nuclear spectroscopy measurements, as they cannot be excluded by a veto in the time domain. Instead, periodic annealing of the crystal, e.g.~once a day, has proven to reduce the emission of uncorrelated photons by a factor approaching 100 compared to a crystal stored dark and cold for months. After annealing, the emission increases with an initial rate of $2 \times 10^{-3}$ photons/($\mathrm{s} \times \mathrm{Bq}$) at room temperature. Note also that the onset of the delayed-photon spectrum appears about 40\,nm towards longer wavelengths compared to the prompt-photon spectrum. This allows for more convenient spectral filtering with respect to the sought-after nuclear emission around 160\,nm.

\section{Conclusion}

We have measured the photoluminescence spectrum of CaF$_2$ upon irradiation with VUV light and, for the first time, resolved the substructure of the prominent STE feature. Spectral filtering will allow to remove the luminescence background, which extends down to 260\,nm, from the nuclear isomer signal expected around 160\,nm.

We have developed a consistent understanding of the radioluminescence emission, which comprises two components:~intense flashes of $\mu$s duration following a radioactive decay in the crystal, as well as a constant background of single uncorrelated photons emitted by very long-lived crystal defects. The spectra of these two components are strikingly different with lower cut-offs at 220 and 260\,nm, respectively. The number of ``prompt'' photons can be reduced by a factor of 10 through a mild temperature increase. Additionally, an auxiliary detector, sensitive around 300\,nm, can be used to identify such events and thus provide the gating of a primary detector with peak sensitivity at the expected nuclear signal. The background of ``delayed'' photons can be removed almost entirely by periodic heating of the crystal.

These findings can be used to optimize the protocol of future searches for the nuclear VUV photon using synchrotron radiation, and guide the selection of suitable photo detectors and optical filters.

Future work will include an absolute photoluminescence measurement using well-defined synchrotron radiation. Understanding the reason for the large photoluminescence level in our home-grown crystals will allow us to improve the crystal growing process in this regard. Considering MgF$_2$ as a viable alternative to CaF$_2$, we will assess the doping efficiency of Th into MgF$_2$.

\section{Acknowledgements}

We thank C.~Tscherne and B.~Ullmann for early experimental work, and we thank J.~Sterba und V.~Rosecker for supporting radiochemistry work. The in-house crystal growing was performed in a highly acknowledged collaboration with J.~Friedrich, P.~Berwian, and K.~Semmelroth from Fraunhofer IISB (Erlangen, Germany). We are indebted to R.~Uecker and R.~Bertram from IKZ Berlin for providing us with a $^{232}$Th:CaF$_2$ crystal. Furthermore, we greatly appreciate fruitful discussions with E.~Peik and G.~Kazakov. This work was supported by the ERC project 258604-NAC and performed within the framework of the newly established nuClock consortium. S.~St.~acknowledges support by the Vienna Center for Quantum Science and Technology (VCQ), and M.~S.~acknowledges support by the Vienna Doctoral Program on Complex Quantum Systems (CoQuS) of the Austrian Science Funds (FWF).

\end{document}